\documentclass[a4paper]{article}

\usepackage{hyperref}
\usepackage{subfigure}
\usepackage{color}
\usepackage{soul}
\usepackage{amsmath}
\usepackage{amssymb}
\usepackage{graphicx}   
\usepackage{bm}
\usepackage{color}
\usepackage{cancel}
\usepackage{slashed}
\usepackage{subfigure}
\usepackage{soul}

\newcommand{\Ket}[1]{\left\vert #1\right\rangle}
\newcommand{\Bra}[1]{\left\langle #1\right\vert}
\newcommand{\PKet}[1]{\vert #1\rangle}

\newcommand{\diss}{{\cal D}}
\newcommand{\gammadiss}{\gamma^{\mathrm{diss}}}
\newcommand{\gammadeph}{\gamma^{\mathrm{deph}}}
\renewcommand{\gammadiss}{\gamma}
\renewcommand{\gammadeph}{\tilde{\gamma}}

\newcommand{\fket}[3]{\Ket{#1}\Ket{#2#3}}
\newcommand{\fproj}[3]{\Ket{#1}\Bra{#1} \otimes \Ket{#2#3}\Bra{#2#3}}

\newcommand{\ii}{\mathrm{i}}
\newcommand{\ee}{\mathrm{e}}

\newcommand{\ignore}[1]{}

\newcommand{\figsize}{0.42}

\renewcommand{\title}[1]{{ \Large\bf \begin{center} #1 \end{center}}}
\newcommand{\authors}[1]{{ \begin{center} #1 \end{center}}}
\newcommand{\address}[1]{{ \it \begin{center} #1 \end{center}}}

\usepackage{geometry}
\begin{document}

\title{Synchronizing two superconducting qubits through a dissipating resonator}

\authors{B. Militello and A. Napoli}

\address{
Universit\`a degli Studi di Palermo, Dipartimento di Fisica e Chimica - Emilio Segr\`e, Via Archirafi 36, 90123 Palermo \\
INFN Sezione di Catania, Via Santa Sofia 64, 95123 Catania, Italy
}

\abstract{
A system consisting of two qubits and a resonator is considered in the presence of different sources of noise, bringing to light the possibility for making the two qubits evolve in a synchronized way. A direct qubit-qubit interaction turns out to be a crucial ingredient as well as dissipation processes involving the resonator. The detrimental role of local dephasing of the qubits is also taken into account.
}



\section{Introduction}

Synchronization of physical systems plays an important role in many fields~\cite{ref:Acebron2005} . It consists of a dynamical alignment of systems, which means that different physical systems, characterized by different natural frequencies, due to some coupling turn out to evolve in such a way to exhibit a common frequency, usually different from the natural ones. The systems can be classical, as for example two clocks or metronomes~\cite{ref:Pantanleone2002,ref:Maianti2009}, people walking on a bridge~\cite{ref:Eckhardt2007} or biological systems~\cite{ref:Angelini2004}, but they can also be quantum. The typical situation involves  two or more interacting quantum oscillators, possibly forming a network, where clusters of synchronized oscillators can be obtained~\cite{ref:Giorgi2012,ref:Manzano2013,ref:Militello2018}. In the realm of quantum mechanics also two-level systems have been considered, as for example the case of two interacting two-state systems undergoing dissipation which dynamically align after a certain time~\cite{ref:Bellomo2017,ref:Cattaneo2021} or two atoms in a cavity coupled through the relevant mode~\cite{ref:Tian2020}. 
In all such cases, the key ingredients for obtaining synchronization are the formation of correlations (typically in the form of entanglement) and dissipation, which drives the system towards suitable superpositions of a limited number of states sharing one frequency or different but very close ones. 
Indeed, generally speaking, persistence of quantum correlations and entanglement in dissipating systems even at equilibrium has extensively been proven~\cite{ref:Vedral,ref:Osterloh,ref:Militello2010a}, but in out-of-equilibrium systems such correlations can induce dynamical alignment. 
Recently, the possibility of obtaining synchronization of two quantum harmonic oscillators coupled to the same dissipating qubits has been predicted~\cite{ref:Militello2017}. Here we want to analyze the complementary situation, consisting of two qubits interacting with a dissipating harmonic oscillator, which are driven to synchronization. In fact, in standard applications, for example in quantum information and technologies, qubits play a fundamental role. Therefore, realizing networks of synchronized qubits which exchange information, instead of the typical networks of oscillators, could be of significant usefulness.

There are several physical scenarios which one can focus on, but superconducting devices are one of the most promising. For example, over the last decades, circuit QED involving superconducting artificial atoms has proven to be a more versatile scenario than the more traditional counterpart, i.e., the cavity QED~\cite{ref:You2011,ref:Xiang2013}. Superconducting circuits offer for example the possibility to reach strong coupling regime where the joint system becomes anharmonic, allowing experiments in nonlinear optics and quantum information at the single photon level~\cite{ref:Wallraff2004}. Superconducting devices can be fabricated using modern integrated circuit technology and their properties, as for example their energies, can be adjusted in situ and determined by circuit parameters, allowing for implementing devices with desired features~\cite{ref:Xiang2013}. These systems thus offer a rich space of parameters and possible operation regimes allowing for the realization of a pletora of Hamiltonian models involving artificial atoms (including qubits as a special case) and resonators~\cite{ref:Blais2007}. 
Moreover, superconducting quantum circuits are privileged candidates for large-scale quantum computing and they  have been used or proposed to implement quantum gates~\cite{ref:Chow2011,ref:Chow2012}, but also to generate entangled states~\cite{ref:DiCarlo2010}, show violation of Bell-type inequalities~\cite{ref:Ansmann2009} or study thermodynamics at a quantum level~\cite{ref:Pekola2015,ref:Cherubim2019,ref:Pekola2019,ref:Elouard2020}.

Similarly to natural atoms, superconducting artificial atoms are plagued by the presence of environments, so that generally one has to take into account both dissipation and decoherence.  Also in this case however we can explore different region of parameters characterized the nonunitary dynamics of the system. Indeed, different values and hierarchies of the relevant dissipation and dephasing rates can be considered, depending for example on whether we are considering transmonic qubits coupled to a waveguide~\cite{ref:Rigetti2012} or charge qubits coupled to a superconductor resonator~\cite{ref:Blais2007}, or other possible configurations\cite{ref:YongLu2021,ref:Sevriuk2019,ref:Jones2013,ref:Lu2021}.
The key tool to take into account environmental effects is a master equation. Generally speaking, in the case of different interacting physical systems the master equation can be obtained with a phenomenological approach, by summing up the dissipators associated to different sources of noise that the single subsystems are subjected to, or by deriving the microscopic master equation starting from the complete system-environment Hamiltonian, including the interaction between the subsystems from the beginning~\cite{ref:MilitelloPRA2007}. In superconducting devices microscopic models for the dephasing mechanisms are available~\cite{ref:Beaudoin2015}, which could be reconsidered in the presence of interaction of the qubits between each other and with the resonator. Generally speaking, in the weak damping limit the predictions coming from phenomenological and microscopic models essentially coincide~\cite{ref:MilitelloPRA2010,ref:ScalaOpts2011}. Therefore, since in this paper we will focus on the weak damping, for the sake of simplicity we will use a phenomenological approach as in Blais {\it et al.}~\cite{ref:Blais2007}.

Dissipation is not necessarily a detrimental occurrence, since in some cases it can even help to obtain interesting physical behaviors, as mentioned before in connection with dynamical alignment. In fact, we show in this paper that a dissipating resonator can drive two qubits interacting with it in a synchronized state of motion.
The material is organized as follows. In section 2 the Hamiltonian model describing the system as well as the master equation governing its dynamics are presented. Section 3 is devoted to the discussion of the appearance of possible protected states and, under appropriate conditions, the occurrence of synchronization phenomena is predicted. Numerical simulations corroborate the theoretical analysis.
The detrimental effects of local qubit dephasing is briefly discussed in Section 4. Then, in the same section, the peculiarity of microcanonical evolutions under pure dephasing is discussed and analyzed.
Finally, in section 5,  we give conclusive remarks.

\section{The Model}

\subsection{Hamiltonian and dissipators}

Our model consists of two qubits with different energy gaps interacting with a quantum harmonic oscillator (resonator) whose frequency is, in general, different from those of the two qubits. We also consider a possible direct interaction between the qubits:
\begin{equation}\label{eq:Hamiltonian1}
H = \sum_{j=A,B} \frac{\Omega_j}{2} \sigma^j_z + \Omega_r a^\dag a +  \sum_{j=A,B} \kappa_j (\sigma_-^j a + \sigma_-^j a^\dag) + \kappa_{AB} \left( \sigma_-^A \sigma_+^B + \sigma_+^A \sigma_-^B \right)\,,
\end{equation}
where $\Omega_A$ and $\Omega_B$ are the natural frequencies of the qubits described by the relevant Pauli operators $\sigma^j_z$,$\sigma^j_x$ and $\sigma^j_y$;  $\Omega_r$ is the natural frequency of the oscillator described by the annihilation and creation operators $a$ and $a^\dag$ and by the number operator $a^\dag a$; $\kappa_A$ and $\kappa_B$ are the qubit-resonator coupling strengths, while $\kappa_{AB}$ is the qubit-qubit coupling strength; both the couplings involve the circular Pauli operators $\sigma^j_\pm = (\sigma_x \pm \ii \sigma_y) / 2$.
The qubit-resonator terms are given in the rotating wave approximation, which is a necessary approximation since it influences the structure of the eigenstates and then, as we will see, the structure of the stable states which are responsible for the synchronization processes. This approximation is valid because of the fact that, in the physical conditions we are going to focus on, the qubit-resonator coupling constant is much smaller than the natural frequencies of the qubits and oscillator.

Since all the parts of the system are interacting with the environment, different sources of noise are present whose effects can be effectively described through a phenomenological master equation involving dissipation and local pure dephasing for the qubits and dissipation for the resonator~\cite{ref:Blais2007}:
\begin{equation}\label{eq:MastEq1}
\dot\rho = -\ii [H, \rho] + \gammadiss_r \diss[a]\rho + \sum_{j=A,B} \gammadiss_j \diss[\sigma^j_-]\rho +  \sum_{j=A,B} \gammadeph_j \diss[\sigma^j_z]\rho \,, 
\end{equation}
where
\begin{equation}
\diss[X] \rho = X \rho X^\dag - \frac{1}{2} \{X^\dag X, \rho\} \,,
\end{equation} 
and $\gammadiss_r$, $\gammadiss_A$ and $\gammadiss_B$ are the decay rates associated to the dissipation processes of the resonator, qubit $A$ and qubit $B$, respectively, while $\gammadeph_j$ ($j=A,B$) are the rates of the local dephasing processes of the two qubits.
We observe that there is no thermal pumping (which would imply terms like $ \diss[a^\dag]$ and $\diss[\sigma^j_+]$), due to the fact that superconduting devices usually operate at low temperature.

Generally speaking, one could wonder whether the presence of interactions between the parts of the system can change the form of the master equation, in the sense that, starting from a microscopic model of interaction between the whole system and the environment, and considering the interactions between the subsystems (each qubit and the resonator), one reaches the so called microscopic master equation~\cite{ref:MilitelloPRA2007}, which is proved to differ from the one obtained evaluating the dissipator before considering the interaction between the subsystems. Nevertheless, since deviations between the two approaches occur in the high-decay regime~\cite{ref:MilitelloPRA2010,ref:ScalaOpts2011}, while we are focusing on the weak-damping limit, for the sake of simplicity we will use the master equation of \eqref{eq:MastEq1}.

\subsection{Conservation of the excitation number}

An important property of the Hamiltonian in \eqref{eq:Hamiltonian1} is that the total number of excitations,
\begin{equation}
\hat{N} = a^\dag a + \sigma_z^A + \sigma_z^B + 2 \,,
\end{equation}
is a constant of motion, due to the fact that both the free terms and the interaction terms in the rotating wave approximation conserve it. Here the constant $2$ allows for having only positive eigenvalules.

This fact implies a block structure for the Hamiltonian involving quadruplets, when the number of excitations is larger than unity:
\begin{equation}
\fket{n}{-} {-}\,, \,\,\, \fket{n-1}{+} {-}\,, \,\,\, \fket{n-1}{-} {+}\,, \,\,\, \fket{n-2}{+} {+}\,, \qquad n\ge 2\,.
\end{equation}
Then, there is also the triplet with one-excitation states,
\begin{equation}
\fket{1}{-} {-}\,, \,\,\, \fket{0}{+} {-}\,, \,\,\, \fket{0}{-} {+}\,,
\end{equation}
and the singlet, given by the sole ground state $\Ket{G}\equiv\fket{0}{-} {-}$, whose energy is $E_G = -(\Omega_A + \Omega_B)/2$.

In the presence of dephasing, the number operator is still conserved, since the operators $\sigma^j_z$ do commute with $\hat{N}$. On the contrary, 
when dissipation is present, whether involving the qubits, the oscillator or both, subspaces with different numbers of excitations are incoherently coupled and the number operator is not conserved.

\section{Protected states and qubit synchronization}

\subsection{Theoretical analysis}

Let us focus on the case $\gamma_A=\gamma_B=\tilde\gamma_A=\tilde\gamma_B=0$ and $\gamma_r\not=0$. Therefore we have two two-state systems (the qubits) interacting with a dissipating oscillator (the resonator). Recently, it has been analyzed the complementary situation consisting of two oscillators coupled with a dissipating two-state system~\cite{ref:Militello2017}, and it has been brought to the light that, under suitable hypotheses, the two oscillators reach an almost stationary regime where they oscillate at the same frequency, thus leading to a synchronization of them. The reason for this occurrence is that there is the possibility to define two new modes (not necessarily the normal modes), one of which is coupled to the two-state system, then \lq indirectly\rq\, undergoing dissipation, while the second one is decoupled and then insensitive to dissipation. After a long time, the dissipating mode loses all its energy, while the other one persists, imposing to the two oscillators to move with a common frequency.

We will try here to reproduce a similar behaviour, driving the two qubits towards common oscillations corresponding to some protected transitions. 
Basically, we will look for a decoherence-free subspace~\cite{ref:Zanardi1997,ref:Zanardi1998}. In order to be insensitive to the noise, a quantum state should belong to the kernel of the dissipator (which implies it not to have a direct coupling to the environment) and to the kernel of the qubit-oscillator coupling (in order to avoid indirect coupling to the environment), thus obtaining a state which is interaction-free~\cite{ref:Militello2015,ref:Militello2016}  with respect to the qubit-resonator coupling. By imposing the second condition,
\begin{equation}
\left[a \left( \kappa_A\sigma_+^A + \kappa_B\sigma_+^B \right) + a^\dag \left( \kappa_A\sigma_-^A + \kappa_B\sigma_-^B \right)\right]\Ket{P} = 0\,,
\end{equation}
we find 
\begin{equation}
\Ket{P} = \cos\theta \fket{0}{+}{-} + \sin\theta \fket{0}{-}{+}\,, \qquad \tan\theta = - \kappa_A / \kappa_B \,,
\end{equation}
as the only possible solution. This state (i.e., the corresponding projector) belongs to the kernel of the dissipator, since it factorizes the ground state $\Ket{0}$. 
However, this is still not enough to prevent waste of energy from the state $\Ket{P}$. Indeed, if it is coupled through the other Hamiltonian terms (even free terms) to other states which are noise-sensitive, it can effectively decay. Therefore we tune the strength of the qubit-qubit interaction to a specific value which allows for $\Ket{P}$ to be an eigenstate of the Hamiltonian.
We in particular require that $H \Ket{P} = E_P \Ket{P}$, with $E_P$ to be determined. Since $H \Ket{P} =  [(-\kappa_B (\Omega_A - \Omega_B)/2+\kappa_A\kappa_{AB}) \fket{0}{+}{-}+ (\kappa_A (\Omega_B - \Omega_A)/2-\kappa_B\kappa_{AB}) \fket{0}{-}{+}]/(\kappa_A^2+\kappa_B^2)^{1/2}$, we obtain:
\begin{equation}\label{eq:kABValue}
\kappa_{AB} = \frac{(\Omega_A-\Omega_B) \kappa_A\kappa_B}{\kappa_A^2-\kappa_B^2} \,,
\end{equation}
\begin{equation}
E_P =  \frac{(\Omega_B-\Omega_A)(\kappa_A^2+\kappa_B^2)}{2(\kappa_A^2-\kappa_B^2)} \, .
\end{equation}

By tuning the qubit-qubit coupling constant $\kappa_{AB}$ to the value of \eqref{eq:kABValue}, we get that the state $\Ket{P}$ turns out to be protected by noise. There is another noise-insensitive state, which is the ground state $\Ket{G}$. Therefore, after a long time, the only two surviving states are $\Ket{P}$ and $\Ket{G}$, and the system evolution is characterized by a single frequency, which is the $\Ket{P} - \Ket{G}$ transition frequency.
It is interesting to note that under the condition $\kappa_B/\kappa_A = \sqrt{\Omega_B/\Omega_A}$, the energies $E_P$ and $E_G$ turn out to be equal, which implies the absence of oscillations in the long-time regime.

It is the case to observe that when $\kappa_A=\kappa_B$ the eigenvalue equation can be satisfied for any value of $\kappa_{AB}$, provided $\Omega_A=\Omega_B$. But this is a trivial case we are no interested in, since the two qubits would be synchronized from the beginning, having the same natural frequencies. Finally, we emphasize that there is no solution for $\kappa_A=\kappa_B$ and $\Omega_A\not=\Omega_B$.

\subsection{Simulations}

Our theoretical analysis is supported by numerical calculations. Since the master equation describing our system is time-independent, its numerical resolution can be easily performed through the evaluation of the exponential of the matrix representing the master equation, multiplied by the time $t$.
Concerning the parameters, we have considered typical values for the natural frequencies and the coupling constants~\cite{ref:You2011,ref:Xiang2013,ref:Blais2007}. The first ones should lie in the range $5-15$ GHz, while the qubit-resonator coupling strenght should lie in the range $10-200$ MHz. Compatibly, (on a $\kappa_A=100$ MHz basis) we will mainly consider $\Omega_A / \kappa_A = 55$, $\Omega_B / \kappa_A = 70$, $\Omega_r / \kappa_A = 64$, while $\kappa_B$ will be of the same order of $\kappa_A$.
Regarding the decay and dephasing rates, the local ones related to the qubits can be made pretty small, having $0.02$ MHz for local qubit dissipation and $0.3$ MHz for local qubits dephasing, compatible with $\gammadiss_j / \kappa_A \sim 0.0002$ and $\gammadeph_j / \kappa_A \sim 0.003$.
The resonator decay rate can be made even smaller than the qubit counterparts. Nevertheless, since we want to use the dissipation process of the resonator to induce qubit synchronization, we will require the use of a less protected resonator (which is clearly always in the grasp of experimentalists) in order to have $\gammadiss_r$ larger than the other rates. In most simulations we will assume $\gammadiss_r/\kappa_A = 0.5$.

\begin{figure}
\begin{tabular}{cc}
\subfigure[]{\includegraphics[width=\figsize\textwidth]{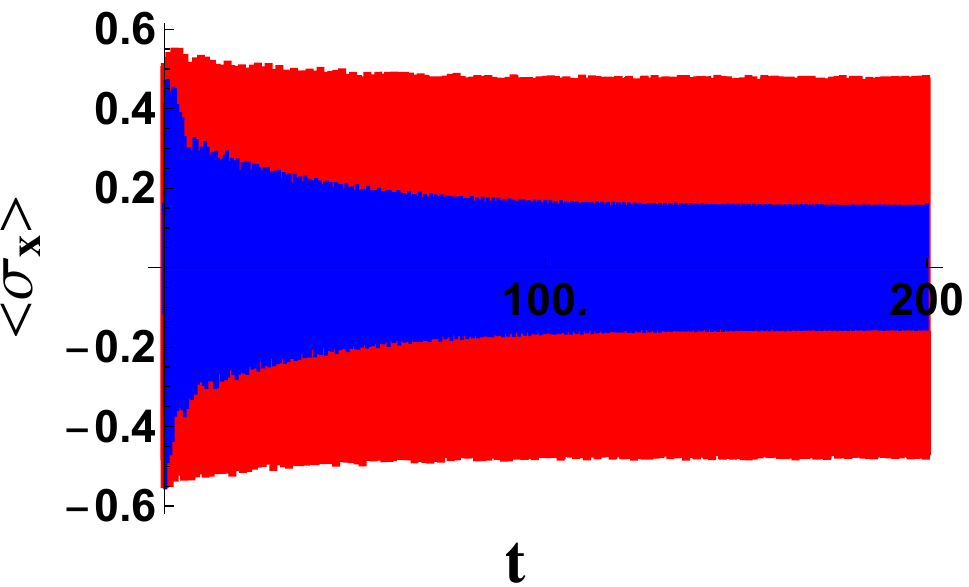}}  &  \subfigure[]{\includegraphics[width=\figsize\textwidth]{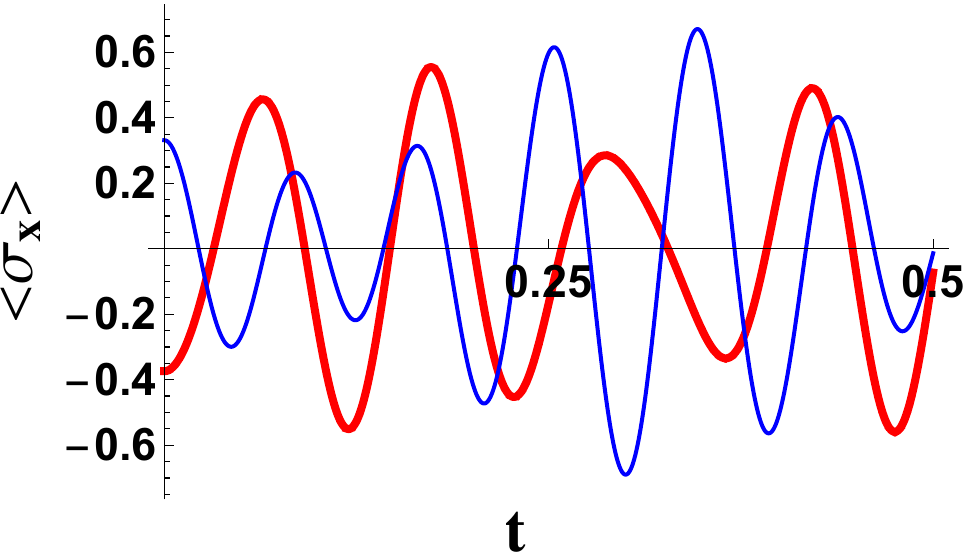}}  \\
\subfigure[]{\includegraphics[width=\figsize\textwidth]{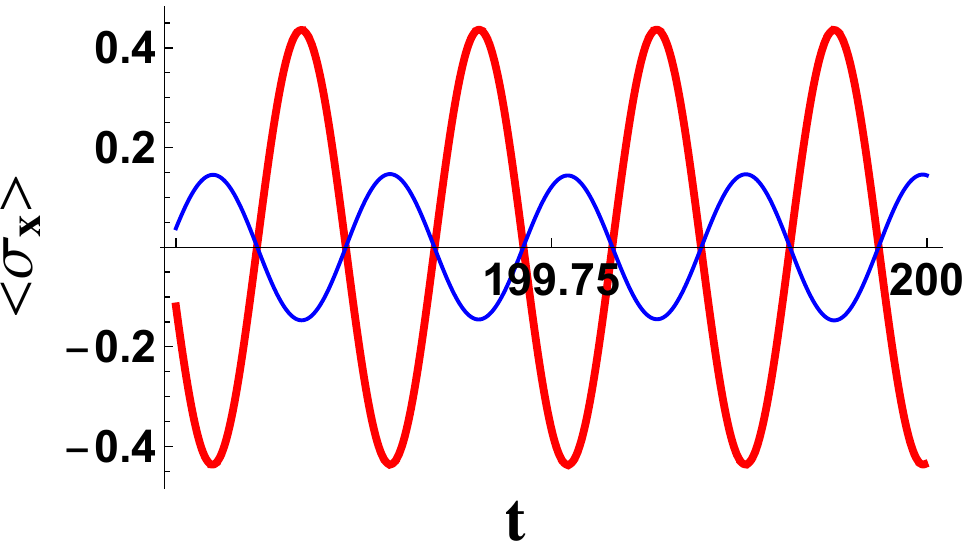}}  &  \subfigure[]{\includegraphics[width=\figsize\textwidth]{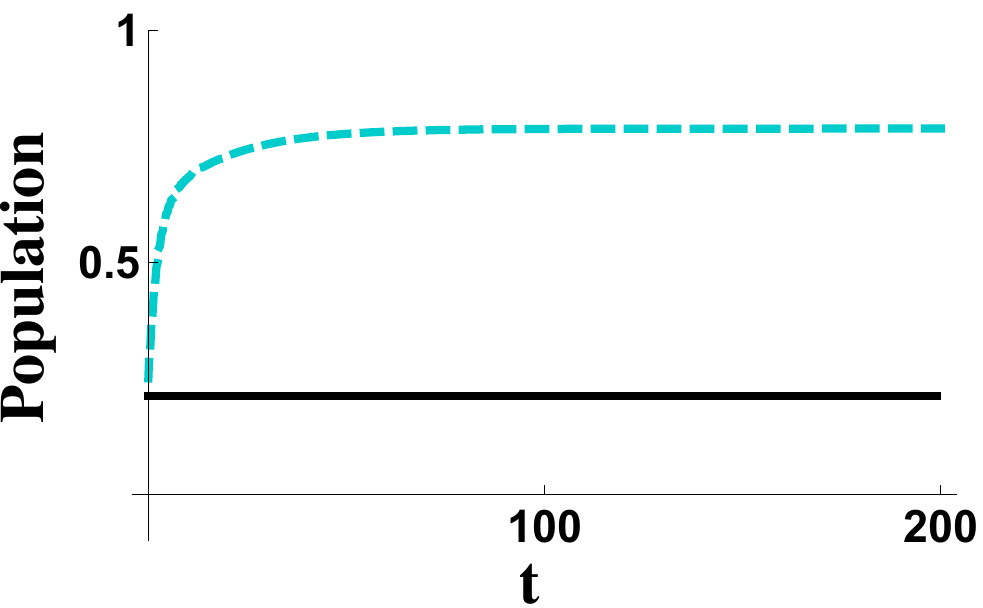}}  
\end{tabular}
\caption{Time evolution of the mean values $\langle \sigma_x^A\rangle$ (bold red line) and $\langle \sigma_x^B\rangle$ (thin blue line) in different time windows, and of the populations of the states $\Ket{P}$ (solid black line) and $\Ket{G}$ (cyan dashed line) in the wide range. Time is given in units of $\kappa_A^{-1}$. The initial state of the system is $\Ket{\psi(0)} = (\Ket{P}+\Ket{G})/2 + \fket{1}{-}{-}/\sqrt{2}$.
The parameters are (in units of $\kappa_A$): $\Omega_A = 55$,  $\Omega_B = 70  $,  $\Omega_r =  64 $, $\kappa_B = 3$,  $\gamma_r =  0.5$, $\gammadiss_A = \gammadiss_B = \gammadeph_A = \gammadeph_B =  0$. The coupling $\kappa_{AB}$ is determined by the decoupling condition in \eqref{eq:kABValue}.}\label{fig:synchro_PGD_id}
\end{figure}

\begin{figure}
\begin{tabular}{cc}
\subfigure[]{\includegraphics[width=\figsize\textwidth]{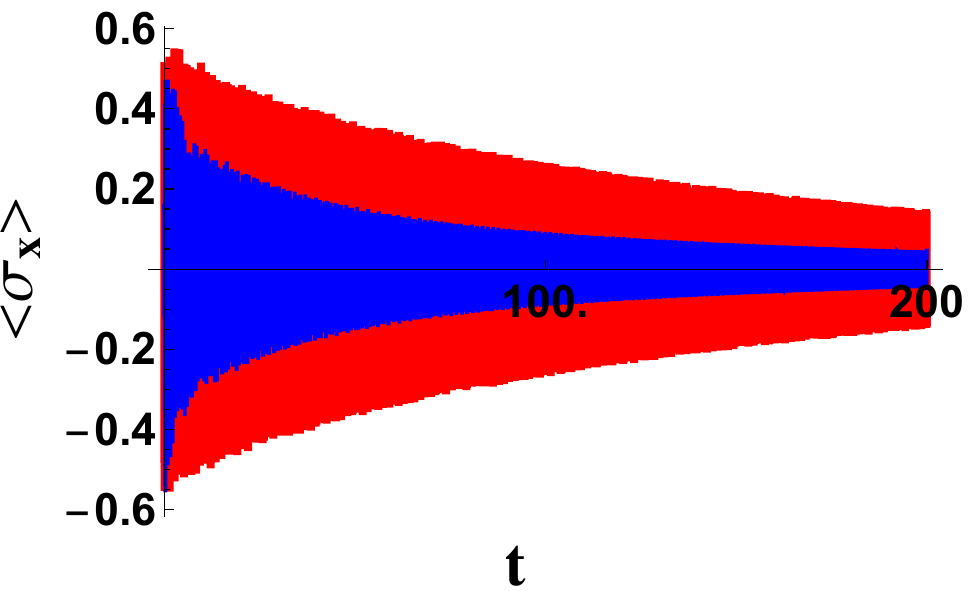}}  &  \subfigure[]{\includegraphics[width=\figsize\textwidth]{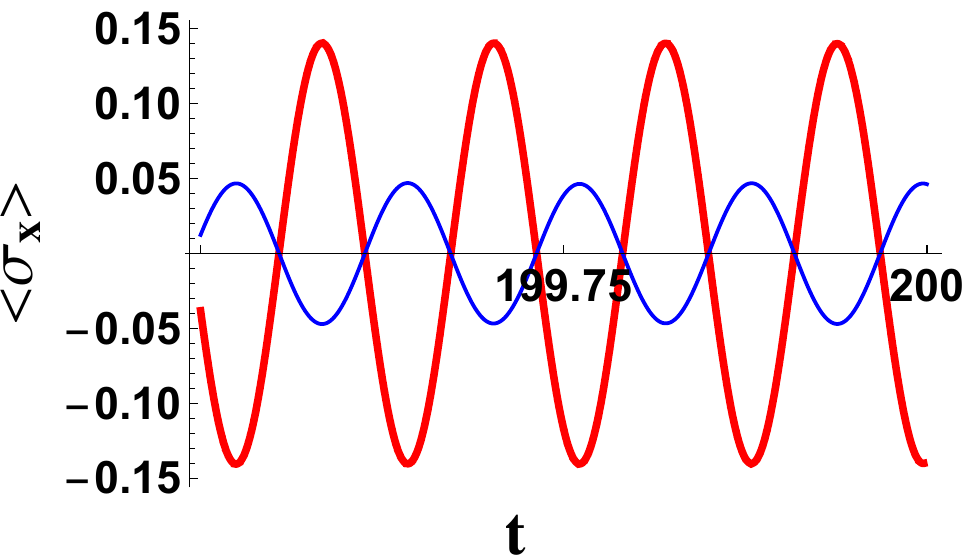}}  \\
\end{tabular}
\caption{The same initial condition and parameters of fig.~\ref{fig:synchro_PGD_id}, except for the local qubit dephasing and dissipation rates: $\gammadiss_A = \gammadiss_B = 0.0002$, and $\gammadeph_A = \gammadeph_B =  0.003$}\label{fig:synchro_PGD_deph}
\end{figure}

In fig.~\ref{fig:synchro_PGD_id} we report the appearance of a synchronized evolution of the two qubits, singled out by the time evolution of the two expectation values $\langle \hat{\sigma}^j_x\rangle$, $j=A,B$, and explained in terms of the populations of the two stationary states $\Ket{P}$ and $\Ket{G}$. 
The system is assumed to be prepared in the state $\Ket{\psi(0)}=(\Ket{P}+\Ket{G}+\sqrt{2}\fket{1}{-}{-})/2$, which is an equal-weight superposition of the two protected states \lq soiled\rq\, by a non-protected one. We assume no local qubits dissipation or dephasing.
The wide-range plot (\ref{fig:synchro_PGD_id}a) shows a general loss of energy of the system associated with a diminishing of the amplitude of $\hat{\sigma}^B_x$. The short-time plot (fig.~\ref{fig:synchro_PGD_id}b) shows that the two signals are very different in the very first part of the evolution, while after a long time, as shown in fig.~\ref{fig:synchro_PGD_id}c, the two expectation values become two clean sinusoidal signals with the same frequency. From fig.\ref{fig:synchro_PGD_id}d we infer that the dynamical alignment of the two qubits is concomitant with the increase of population of the ground state, to the point where only $\Ket{P}$ and $\Ket{G}$ are present in the state of the system. In fig.~\ref{fig:synchro_PGD_deph} we consider the effects of the local dissipation and dephasing of the qubits. It is well visible that the amplitudes of the oscillations are smaller than in the previous case, but not dramatically smaller, making the effect still appreciable.

\begin{figure}
\begin{tabular}{cc}
\subfigure[]{\includegraphics[width=\figsize\textwidth]{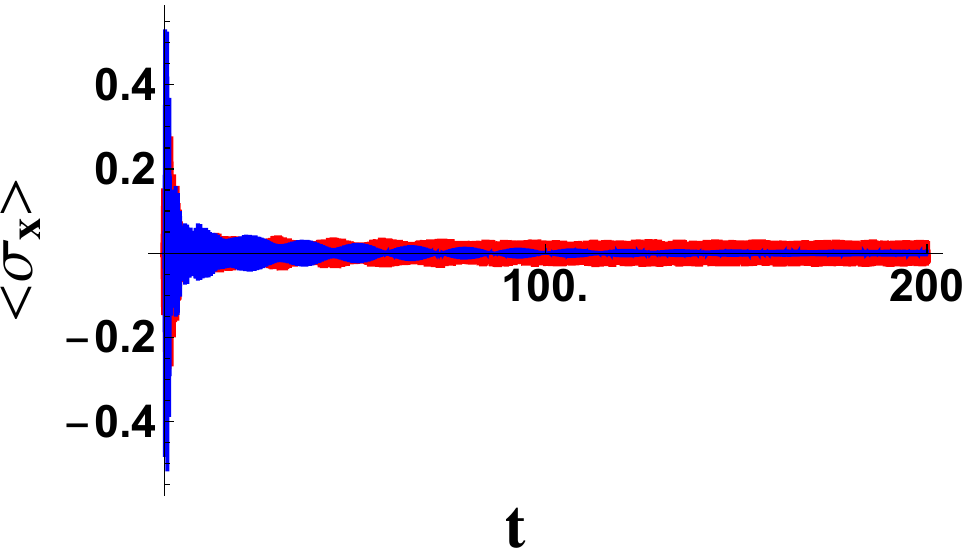}}  &  \subfigure[]{\includegraphics[width=\figsize\textwidth]{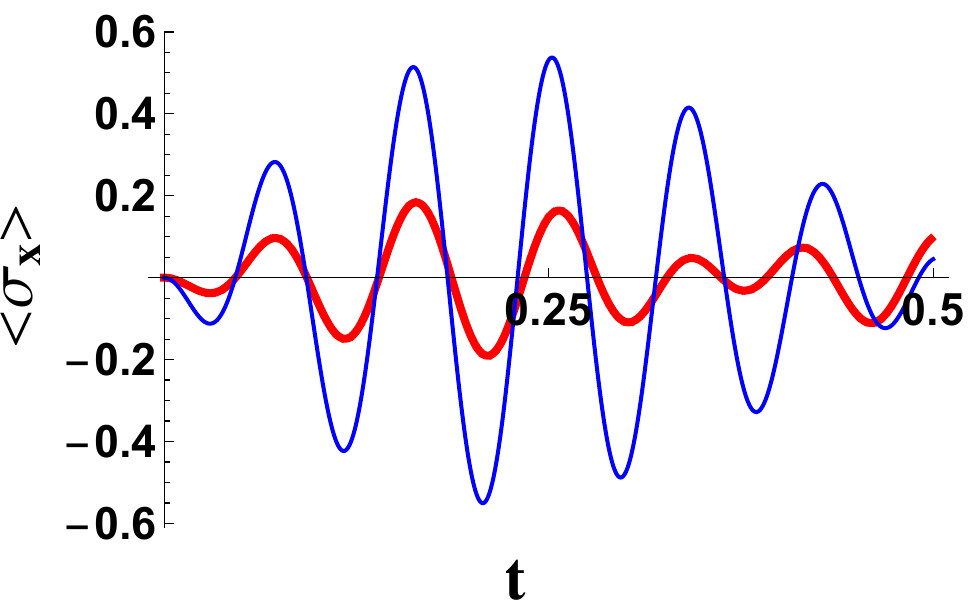}}  \\
\subfigure[]{\includegraphics[width=\figsize\textwidth]{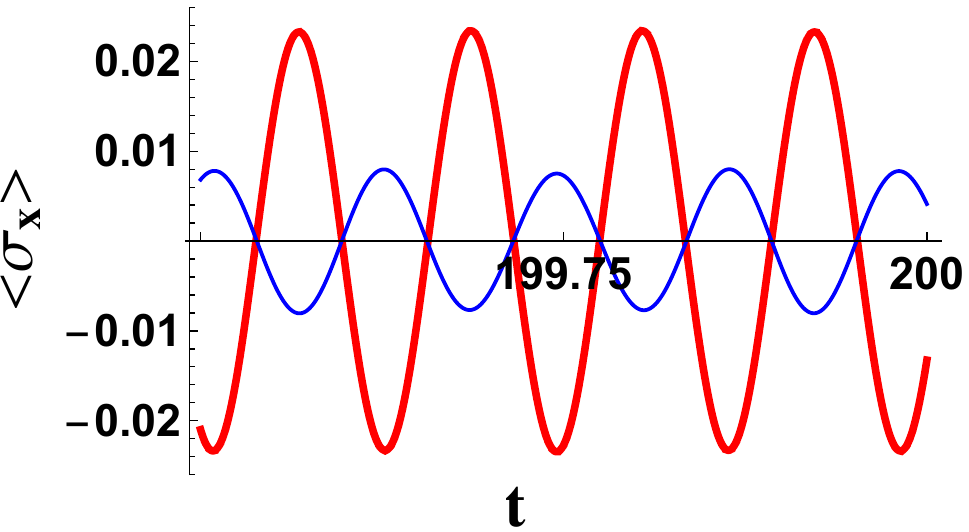}}  &  \subfigure[]{\includegraphics[width=\figsize\textwidth]{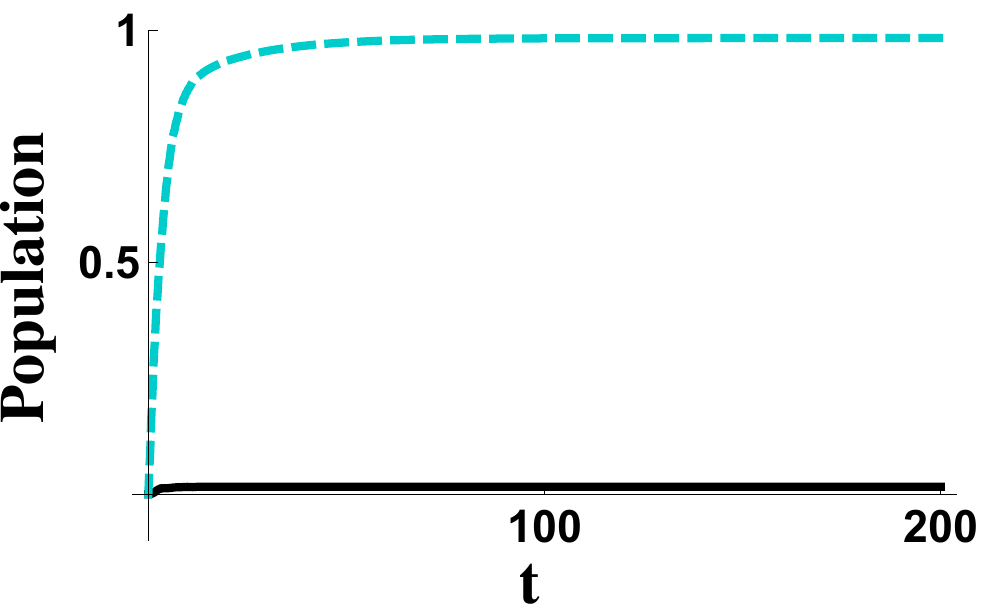}}  
\end{tabular}
\caption{Time evolution of the mean values $\langle \sigma_x^A\rangle$ (bold red line) and $\langle \sigma_x^B\rangle$ (thin blue line) in different time windows, and of the populations of the states $\Ket{P}$ (solid black line) and $\Ket{G}$ (cyan dashed line) in the wide range. Time is given in units of $\kappa_A^{-1}$. The initial state of the system is $\Ket{\psi(0)} = (\fket{2}{-}{-} + \fket{1}{-}{-})/\sqrt{2}$.
The parameters are (in units of $\kappa_A$): $\Omega_A = 55 $,  $\Omega_B = 70$,  $\Omega_r =  64 $,  $\kappa_B = 3$,  $\gamma_r =  0.5$,
$\gamma_A = \gamma_B = \tilde\gamma_A = \tilde\gamma_B =  0$. The coupling $\kappa_{AB}$ is determined by the decoupling condition in \eqref{eq:kABValue}.}\label{fig:synchro_SPR}
\end{figure}

In fig.~\ref{fig:synchro_SPR} we analyze a situation similar to that of fig.~\ref{fig:synchro_PGD_id}, but in connection with a different initial condition: $\Ket{\psi(0)} = (\fket{2}{-}{-} + \fket{1}{-}{-})/\sqrt{2}$. Since the amount of coherence $\Bra{P}\rho(t)\Ket{G}$ (with $\rho(t)$ the state of the system at time $t$) at $t=0$ is very small, its final value turns out to be small even at long time, which turns out into oscillations with very small amplitudes.
It is also well visible that the final population of the state $\Ket{P}$ is pretty small, though not vanishing.

\begin{figure}
\begin{tabular}{cc}
\subfigure[]{\includegraphics[width=\figsize\textwidth]{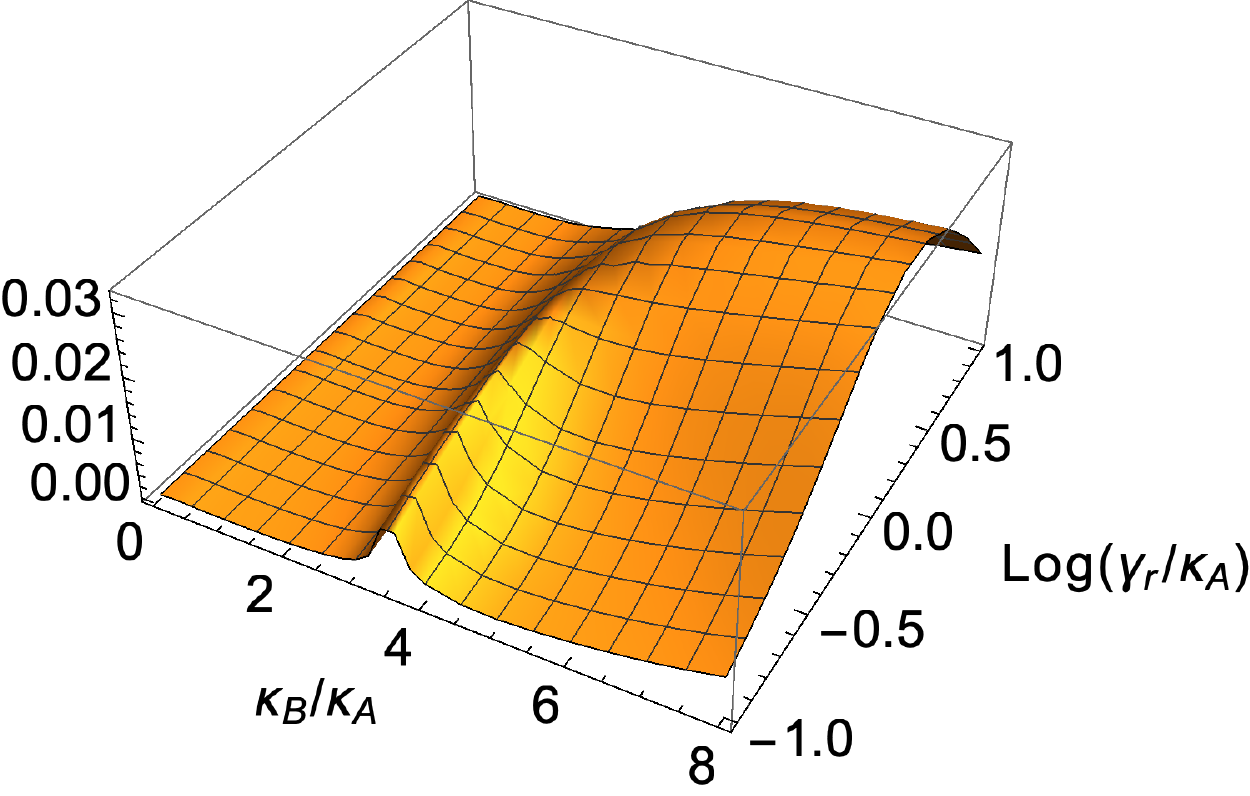}}  &  \subfigure[]{\includegraphics[width=\figsize\textwidth]{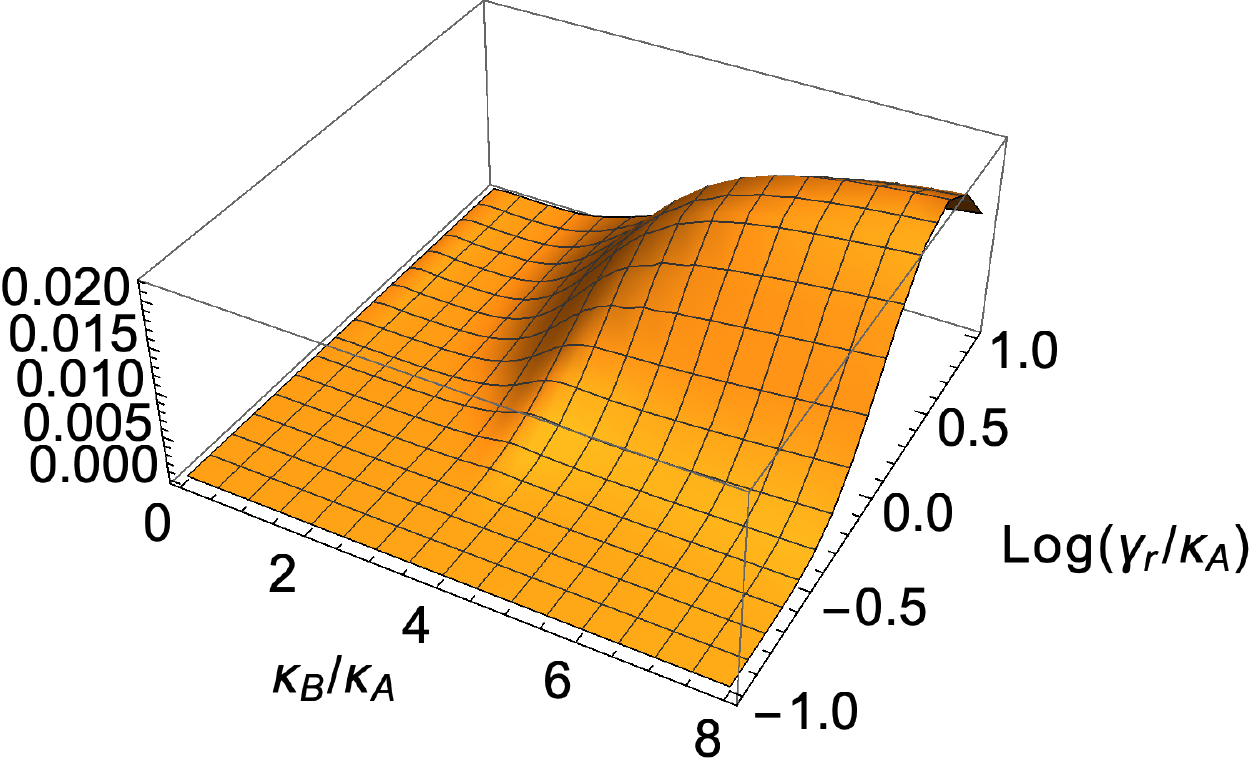}} 
\end{tabular}
\caption{Modulus of the coherence $\Bra{P}\rho(t)\Ket{G}$ at time $t = 200 / \gamma_r$ as function of $\kappa_B/\kappa_A$ and $\log(\gamma_r/\kappa_A)$, for the initial state $\fket{\alpha}{-}{-}$, with $\alpha=1$, in the absence of dephasing (a) and in the presence of dephasing (b).  
The relevant parameters are (in units of $\kappa_A$): $\Omega_A = 55  $, $\Omega_B = 70  $, $\Omega_r =  64 $, $\gammadiss_A=\gammadiss_B=0$, and $\kappa_{AB}$ is determined by the decoupling condition in \eqref{eq:kABValue}. In (a) we have $\gammadiss_A=\gammadiss_B=\gammadeph_A=\gammadeph_B=0$, while in (b) $\gammadiss_A=\gammadiss_B=0.0002$ and $\gammadeph_A=\gammadeph_B=0.003$.}\label{fig:extended_kb_gr}
\end{figure}

In fig.~\ref{fig:extended_kb_gr} we plot the long-time coherence as a function of $\kappa_B$ and $\gammadiss_r$ (here forcing the model to the strong-damping limit, since we reach the value $\gammadiss_r=10\kappa_A$), for two different initial conditions. In fig.~\ref{fig:extended_kb_gr}a a coherent state is considered for the resonator, while the two qubits are in their natural ground state: $\Ket{\psi(0)}=\fket{\alpha}{-}{-}$, with $\alpha=1$; in fig.~\ref{fig:extended_kb_gr}b it is reported the same situation but in the presence of local qubit dissipation and dephasing.

\section{Dephasing}

The appearance of synchronized motion of the two qubits is based on the interaction between the two-state systems and the dissipating resonator, which drives most of the two-qubit states towards the ground state, leaving only a specific two-state superposition involving only a transition frequency. 
Local qubit dissipation and dephasing processes instead tend to destroy synchronization, as it is well visible, for example in fig.~\ref{fig:synchro_PGD_deph} and \ref{fig:extended_kb_gr}, where small qubit

Pure dephasing itself is an interesting phenomenon from the thermodynamical point of view, since it is the main mechanism of relaxation of a system into a microcanonical state.
Usually, dephasing does not alter energy of the system and induce a random distribution of the states with equal energies. Since in our case the dephasing mechanism does not conserve the energy of the system but only its total number of excitations, it is interesting to observe the structure of the equilibrium state in such a situation.
Let us then focus on the pure dephasing (thus assuming $\gammadiss_r=\gammadiss_A=\gammadiss_B=0$):
\begin{equation}
\dot\rho = -\ii [H, \rho] + \sum_j \gammadeph_j ( \sigma_z^j \rho \sigma_z^j - \rho ) \,.
\end{equation}

The microcanonical state related to a subspace with fixed number of excitations is a stationary state: $\rho = g_n^{-1} \hat{\Pi}_n$ \, $\Rightarrow$  \, $-\ii [H, \rho] +  \sum_j \gammadeph_j ( \sigma_z^j \rho \sigma_z^j - \rho ) = 0$ \,, where $\hat{\Pi}_n$ denotes the projector onto the subspace with $n$ excitations and $g_n$ is the relevant degeneracy.

It is possible to prove that this is the only possible stationary state, unless specific conditions (given below) are satisfied. Indeed, assume $\rho_n = \sum_k p_{nk} \Ket{\psi_{nk}}\Bra{\psi_{nk}}$, where $\sum_k p_{nk}=1$ and $\Ket{\psi_{nk}}$'s form an orthonormal basis of the subspace with $n$ excitations. The action of the Lindbladian gives:
\begin{equation}
-\ii (\sum_k p_{nk} \Ket{\psi_{nk}}\Bra{\phi_{nk}} - \sum_k p_{nk} \Ket{\phi_{nk}}\Bra{\psi_{nk}}) + \sum_j \gammadeph_j \left[ 
\sum_k p_{nk} \sigma^j_z\Ket{\psi_{nk}}\Bra{\psi_{nk}}\sigma^j_z - \sum_k p_{nk} \Ket{\psi_{nk}}\Bra{\psi_{nk}}
\right] \,,
\end{equation}
where $\Ket{\phi_{nk}} = H\Ket{\psi_{nk}}$ are non normalized states.
In order to have this quantity to vanish, the first two sums must give zero, which implies $\Ket{\phi_{nk}} \propto \Ket{\psi_{nk}}$, i.e., the states $\Ket{\psi_{nk}}$ are Hamiltonian eigenstates. In order to obtain zero the other two terms should compensate each other. Since the states  $\Ket{\psi_{nk}}$ are orthogonal, the only possibility is that, for every $k$, \, $\sigma^j_z\Ket{\psi_{nk}} = \ee^{\ii\varphi_{nkj}} \PKet{\psi_{nm_j}}$ provided $p_{nk}=p_{nm}$. Of course, in such a case one also has $\sigma^j_z\Ket{\psi_{nm}} = \ee^{-\ii\varphi_{nkj}} \PKet{\psi_{nk_j}}$. Therefore, unless specific eigenstates of the Hamiltonian exist which are eigenstates of both the $\sigma^j_z$ or are mapped by such operators to other Hamiltonian eigenstates, there is no stationary state different from the microcanonical one.

\begin{figure}
\begin{tabular}{cc}
\subfigure[]{\includegraphics[width=\figsize\textwidth]{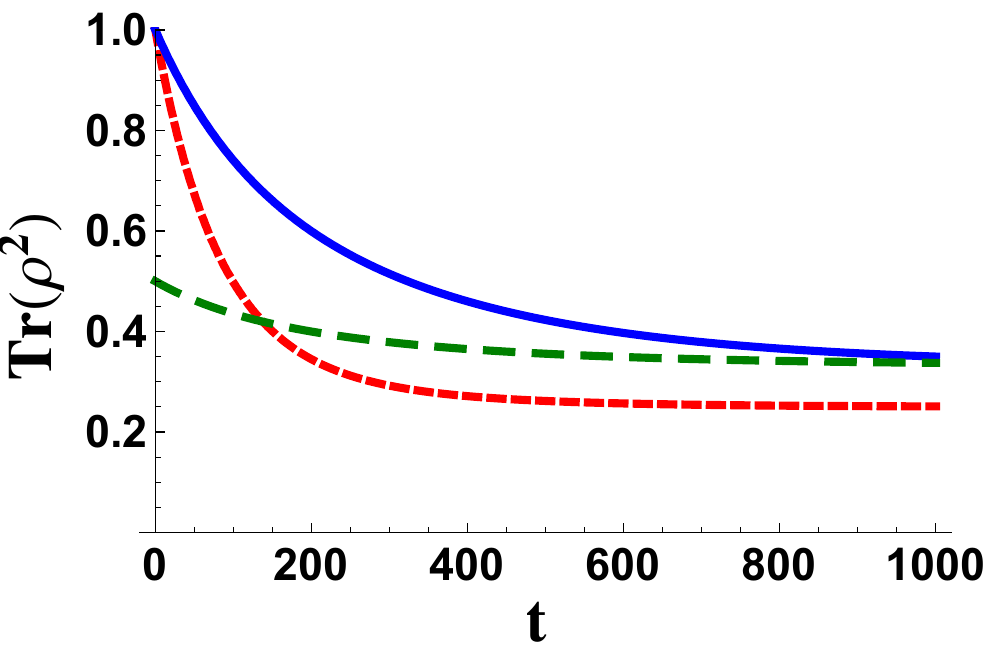}}  &  \subfigure[]{\includegraphics[width=\figsize\textwidth]{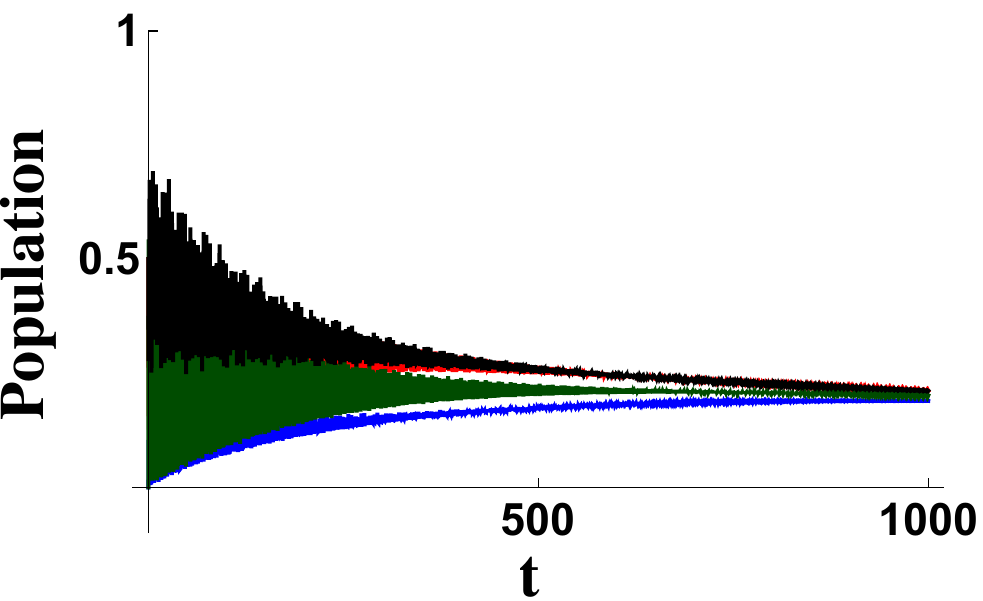}}  \\
\end{tabular}
\caption{
In (a) we report the time evolution of the linear entropy for different initial states: $\fket{1}{-}{-}$ (blue solid line), $\fket{2}{-}{-}$ (red dotted line) and the mixed state $(\fproj{0}{-}{+}+\fproj{0}{+}{-})/2$ (green dashed line). 
In (b) the time evolutions of the populations of the states $\fket{2}{-}{-}$ (red line),  $\fket{1}{+}{-}$ (blue line),  $\fket{1}{-}{+}$ (green line),  $\fket{0}{+}{+}$ (black line) are plotted, when the system is prepared in  $\Ket{\psi(0)} = (\fket{0}{+}{+}+ \fket{2}{-}{-})/\sqrt{2}$. 
Time is given in units of $\kappa_A^{-1}$. The coupling $\kappa_{AB}$ is determined by the decoupling condition, while the other parameters are (in units of $\kappa_A$): $\Omega_A = 55  $, $\Omega_B = 70  $, $\Omega_r =  64 $, $\kappa_B = 3$, $\gammadeph_A = \gammadeph_B=0.003$, $\gammadiss_r=\gammadiss_A=\gammadiss_B=0$.}\label{fig:microcanonical}
\end{figure}

On the basis of the previous results, one can expect that when the system is prepared into an eigenstate of the number operator $\hat{N}$ (i.e., every state belonging to a multiplet with fixed $n$), the evolution naturally brings the system toward the microcanonical state, which is characterized by a maximum von Neumann entropy (${\cal S}(\rho)\equiv - \mathrm{Tr}(\rho\log\rho)$ $\rightarrow$ ${\cal S}(g_n^{-1}\hat{\Pi}_n)=\log g_n$) and a minimum linear entropy (${\cal P}(\rho) \equiv \mathrm{Tr}\rho^2$ $\rightarrow$ ${\cal P}(g_n^{-1}\hat{\Pi}_n)=g_n^{-1}$). In fig.~\ref{fig:microcanonical}a we show the time behavior of the linear entropy when the system is prepared in different states. The two initial conditions belonging to the triplet with $n=1$ lead to an asymptotic purity of $1/3$, while the initial state belonging to the quadruplet with $n=2$ evolves in such a way to give a purity equal to $1/4$, as expected.
In fig.~\ref{fig:microcanonical}b an example of evolution of the populations of states belonging to the quadruplet with $n=2$ is given, assuming that the system is prepared in a state with two excitations. As it is well visible, all populations eventually reach the value 1/4.

The peculiarity of the microcanonical state the system approaches is that it does not describe a situation where all the eigenstates of the Hamiltonian with the same energy have equal probability. Since the given constraint relates to the number of excitations, the uniform probability distribution refers to the states with the same number of excitations.

\section{Discussion and Conclusions}

In this paper we have studied a pseudo-Dicke model describing the interaction between a harmonic oscillator (the resonator) and two two-state systems (qubits) which can be realized with superconducting devices. The qubit-resonator is assumed to be in the rotating wave approximation, which allows for the conservation of the total number of excitations. Besides this interaction , a direct qubit-qubit coupling is considered. Moreover, since all the components are subjected to an interaction with the environment, dephasing and dissipation processes are also included in the model.

We have seen that it is possible to identify a state which is insensitive to the qubit-resonator interaction and that, by suitably tuning the qubit-qubit coupling, such a state can be made an eigenstate of the Hamiltonian, hence allowing for the relevant qubit state to be stationary. This state and the ground state of the system form a protected subspace. Therefore, after a large enough time, the system inevitably relaxes toward a mixture involving these two states with a possible residual coherence. As a consequence, the two-qubit system exhibits oscillations at the frequency which separates the protected and the ground states. This leads to an effective synchronization, which provides the complementary scenario to that analyzed in Ref.~\cite{ref:Militello2017} where two oscillators synchronize due to the common interaction with a dissipating two-state system. The direct coupling between the qubits play a crucial role, since it allows for avoiding transitions from the state which is insensitive to the qubit-resonator coupling towards states which are affected by that interaction term.

We have also analyzed the role of local qubit dissipation and dephasing processes, showing that they are able to damage the processes inducing synchronization. On the other hand, dephasing itself makes the system relax towards a pseudo-microcanonical state, which is characterized by equiprobability of the states with the same number of excitations.
This analysis shows how rich can be the dynamics which can be obtained with superconducting devices.




\begin{thebibliography}{999}

\bibitem{ref:Acebron2005} J. A. Acebr\'on, L. L. Bonilla, C. J. P\'erez Vicente, F. Ritort, and R. Spigler, The Kuramoto model: A simple paradigm for synchronization phenomena, {\it Rev. Mod. Phys.} {\bf 2005} {\it 77}, 137.

\bibitem{ref:Pantanleone2002} J. Pantaleone, Synchronization of metronomes, {\it Am. J. Phys.} {\bf 2002} {\it 70}, 992.

\bibitem{ref:Maianti2009}  M. Maianti, S. Pagliara, G. Galimberti, and F. Parmigiani, {\it Am. J. Phys.} {\bf 2009} {\it 77}, 834.

\bibitem{ref:Eckhardt2007}  B. Eckhardt, E. Ott, S. H. Strogatz, D. M. Abrams, and A. McRobie, Modeling walker synchronization on the Millennium Bridge, {\it Phys. Rev. E} {\bf 2007} {\it 75}, 021110.

\bibitem{ref:Angelini2004} L. Angelini, G. Lattanzi, R. Maestri, D. Marinazzo, G. Nardulli, L. Nitti, M. Pellicoro, G. D. Pinna, and S. Stramaglia, Phase shifts of synchronized oscillators and the systolic-diastolic blood pressure relation, {\it Phys. Rev. E} {\bf 2004} {\it 69}, 061923.

\bibitem{ref:Giorgi2012} G. L. Giorgi, F. Galve, G. Manzano, P. Colet, and R. Zambrini, Quantum correlations and mutual synchronization, {\it Phys. Rev. A} {\bf 2012}, {\it 85}, 052101.
 
\bibitem{ref:Manzano2013}  G. Manzano, F. Galve, G. L. Giorgi, E. Harn\'andez-Garcia, and R. Zambrini, Synchronization, quantum correlations and entanglement in oscillator networks, {\it Sci. Rep.} {\bf 2013}, {\it 3}, 1439.

\bibitem{ref:Militello2018} B. Militello, D. Chru\'sci\'nski and A. Napoli, Star network synchronization led by strong coupling-induced frequency squeezing, {\it Phys. Scr.} {\bf 2018}, {\it 93}, 025201.

\bibitem{ref:Bellomo2017} B. Bellomo, G. L. Giorgi, G. M. Palma, and R. Zambrini, Quantum synchronization as a local signature of super- and subradiance, {\it Phys. Rev. A}  {\bf 2017}, {\it 95}, 043807.

\bibitem{ref:Cattaneo2021} M. Cattaneo, G. L. Giorgi, S. Maniscalco, G. Sorin Paraoanu, R. Zambrini, Bath-Induced Collective Phenomena on Superconducting Qubits: Synchronization, Subradiance, and Entanglement Generation, {\it Ann. Phys.} {\bf 2021}, {\it 533}, 5, 2100038. 

\bibitem{ref:Tian2020}  Tian-tian Huan, Ri-gui Zhou, Hou Ian, Synchronization of two cavity-coupled qubits measured by entanglement, {\it Scientific Reports} {\bf 2020}, {\it 10},  12975. 

\bibitem{ref:Vedral} M. C. Arnesen, S. Bose, and V. Vedral, Natural Thermal and Magnetic Entanglement in the $1D$ Heisenberg Model, {\it Phys. Rev. Lett.} {\bf 2001}  {\it 87}, 017901.

\bibitem{ref:Osterloh} A. Osterloh, L. Amico, G. Falci, R. Fazio, Scaling of entanglement close to a quantum phase transition, {\it Nature} {\bf 2002}, {\it 416}, 608.

\vskip0cm 

\bibitem{ref:Militello2010a}  B. Militello, A. Messina, Genuine tripartite entanglement in a spin-star network at thermal equilibrium, {\it Phys. Rev. A} {\bf 2011}, {\it 83}, 042305.

\bibitem{ref:Militello2017} B. Militello, H. Nakazato, A. Napoli, Synchronizing quantum harmonic oscillators through two-level systems, {\it Phys. Rev. A} {\bf 2017}, {\it 96}, 023862.

\bibitem{ref:You2011} J. Q. You and Franco Nori, Atomic physics and quantum optics using superconducting circuits, {\it Nature} {\bf 2011} {\it 474}, 589. 

\bibitem{ref:Xiang2013}  Ze-Liang Xiang, Sahel Ashhab, J. Q. You, Franco Nori, Hybrid quantum circuits: Superconducting circuits interacting with other quantum systems, {\it Rev. Mod. Phys} {\bf 2013}, {\it 85}, 623. 
 
\bibitem{ref:Wallraff2004}  A. Wallraff, D. I. Schuster, A. Blais, L. Frunzio, R.- S. Huang, J. Majer, S. Kumar, S. M. Girvin and R. J. Schoelkopf, Strong coupling of a single photon to a superconducting qubit using circuit quantum electrodynamics, {\it Nature} {\bf 2004} {\it 43}, 162-167. 

\bibitem{ref:Blais2007}  A. Blais, J. Gambetta, A. Wallraff, D. I. Schuster, S. M. Girvin, M. H. Devoret, R. J. Schoelkopf, Quantum-information processing with circuit quantum electrodynamics , {\it Phys. Rev. A} {\bf 2007}, {\it 75}, 032329.

\bibitem{ref:Chow2011} J. M. Chow, A. D. C\'orcoles, Jay M. Gambetta, C. Rigetti, B. R. Johnson, John A. Smolin, J. R. Rozen, George A. Keefe, M. B. Rothwell, M. B. Ketchen, and M. Steffen, Simple All-Microwave Entangling Gate for Fixed-Frequency Superconducting Qubits, {\it Phys. Rev. Lett.} {\bf 2011}, {\it 107}, 080502. 

\bibitem{ref:Chow2012} Jerry M. Chow, Jay M. Gambetta, A. D. C\'orcoles, Seth T. Merkel, John A. Smolin, Chad Rigetti, S. Poletto, George A. Keefe, Mary B. Rothwell, J. R. Rozen, Mark B. Ketchen, and M. Steffen, Universal Quantum Gate Set Approaching Fault-Tolerant Thresholds with Superconducting Qubits, {\it Phys. Rev. Lett.} {\bf 2012}, {\it 109}, 060501. 

\bibitem{ref:DiCarlo2010} L. Di Carlo, M. D. Reed, L. Sun, B. R. Johnson, J. M. Chow, J. M. Gambetta, L. Frunzio, S. M. Girvin, M. H. Devoret and R. J. Schoelkopf, Preparation and measurement of three-qubit entanglement in a superconducting circuit, {\it Nature (London)} {\bf 2010}, {\it 467}, 574.  

\bibitem{ref:Ansmann2009} M. Ansmann, H. Wang, Radoslaw C. Bialczak, Max Hofheinz, E. Lucero, M. Neeley, A. D. O'Connell, D. Sank, M. Weides, J. Wenner, A. N. Cleland and J. M. Martinis, Violation of Bell's inequality in Josephson phase qubits, {\it Nature (London)} {\bf 2009}, {\it 461}, 504.

\vskip0cm 

\bibitem{ref:Pekola2015}  Pekola, J. Towards quantum thermodynamics in electronic circuits. {\it Nature Phys} {\bf 2015}, {\it 11}, 118-123.    

\bibitem{ref:Cherubim2019} C. Cherubim, F. Brito and S. Deffner, Non-Thermal Quantum Engine in Transmon Qubits, {\it Entropy} {\bf 2019}, {\it 21}, 545. 

\bibitem{ref:Pekola2019}  J. P. Pekola, I. M. Khaymovich, Thermodynamics in Single-Electron Circuits and Superconducting Qubits, {\it Annu. Rev. Condens. Matter Phys.} {\bf 2019}, {\it 10}, 193.  

\bibitem{ref:Elouard2020} Cyril Elouard, George Thomas, Olivier Maillet, Jukka P. Pekola, Andrew N. Jordan, Quantifying the quantum heat contribution from a driven superconducting circuit, {\it Phys. Rev. E} {\bf 2020}, {\it 102}, 030102.   

\bibitem{ref:Rigetti2012} C. Rigetti, J. M. Gambetta, S. Poletto, B. L. T. Plourde, J. M. Chow, A. D. Corcoles, J. A. Smolin, S. T. Merkel, J. R. Rozen, G. A. Keefe,  M. B. Rothwell, M. B. Ketchen, and M. Steffen, Superconducting qubit in a waveguide cavity with a coherence time approaching 0.1 ms, {\it Phys. Rev. B} {\bf 2012} {\it 86}, 100506(R).  

\bibitem{ref:YongLu2021} Yong Lu,  A. Bengtsson,  J. J. Burnett, E. Wiegand, B. Suri,  P. Krantz, A. Fadavi Roudsari, A. Frisk Kockum, S. Gasparinetti, G. Johansson , P. Delsing, Characterizing decoherence rates of a superconducting qubit by direct microwave scattering, {\it npj Quantum Information} {\bf 2021}, {\it 7}, 35.

\bibitem{ref:Sevriuk2019}  V. A. Sevriuk, K. Y. Tan, E. Hyypp\"{a}, M. Silveri, M. Partanen, M. Jenei, S. Masuda, J. Goetz, V. Vesterinen, L. Gr\"{o}nberg, and M. M\"{o}tt\"{o}nen, Fast control of dissipation in a superconducting resonator, {\it Appl. Phys. Lett.} {\bf 2019}, {\it 115}, 082601. 

\bibitem{ref:Jones2013} P. J. Jones, J. A. M. Huhtama\"{k}i, J. Salmilehto, K. Y. Tan and  M.M\"{o}tt\"{o}nen, Tunable electromagnetic environment for superconducting quantum bits, {\it Sci. Rep.} {\bf 2013}, {\it 3}, 1987. 

\bibitem{ref:Lu2021} Yong Lu, A. Bengtsson, J. J. Burnett, E. Wiegand, B. Suri, P. Krantz, A. Fadavi Roudsari, A. Frisk Kockum, S. Gasparinetti, G. Johansson and P. Delsing, Characterizing decoherence rates of a superconducting qubit by direct microwave scattering, {\it npj Quantum Inf} {\bf 2021}, 7, 35. 

\bibitem{ref:MilitelloPRA2007} M. Scala, B. Militello, A. Messina, J. Piilo, and S. Maniscalco, Microscopic derivation of the Jaynes-Cummings model with cavity losses, {\it Phys. Rev. A} {\bf 2007}, {\it 75}, 013811. 

\bibitem{ref:MilitelloPRA2010} M. Scala, B. Militello, A. Messina and N. V. Vitanov, Stimulated Raman adiabatic passage in an open quantum system: Master equation approach, {\it Phys. Rev. A} {\bf 2010}, {\it 81}, 053847. 

\bibitem{ref:ScalaOpts2011} M. Scala, B. Militello, A. Messina and N. V. Vitanov, Detuning Effects in STIRAP Processes in the Presence of Quantum Noise, {\it Optics and Spectroscopy} {\bf 2011} {\it 111}, 4, 589. 

\bibitem{ref:Beaudoin2015} F. Beaudoin and W. A. Coish, Microscopic models for charge-noise-induced dephasing of solid-state qubits, {\it Phys. Rev. B} {\bf 2015} {\it 91}, 165432.

\bibitem{ref:Zanardi1997}  P. Zanardi and M. Rasetti, Noiseless Quantum Codes, {\it Phys. Rev. Lett.} {\bf 1997}, {\it 79}, 3306.

\bibitem{ref:Zanardi1998}  P. Zanardi and F. Rossi, Quantum Information in Semiconductors: Noiseless Encoding in a Quantum-Dot Array, {\it Phys. Rev. Lett.} {\bf 1998}, {\it 81}, 4752.

\bibitem{ref:Militello2015} D. Chruscinski, A. Messina, B. Militello, and A. Napoli, Interaction-free evolution in the presence of time-dependent Hamiltonians, {\it Phys. Rev. A} {\bf 2015}, {\it 91}, 042123. 

\bibitem{ref:Militello2016} B. Militello, D. Chruscinski, A. Messina, P. Nalezyty, and A. Napoli, Generalized interaction-free evolutions, {\it Phys. Rev. A} {\bf 2016}, {\it 96}, 022113.  



\vskip0cm 



 












\end{thebibliography}

\end{document}